\documentclass[10pt,a4paper]{article}
\usepackage{amsfonts}
\usepackage {graphicx}
\usepackage{amsmath}
\usepackage{amssymb}
\usepackage{amsthm}

\numberwithin{equation}{section}
\numberwithin{figure}{section}

\newtheorem{theorem}{\sc Theorem}[section]
\newtheorem{corollary}[theorem]{\sc Corollary} 
\newtheorem{proposition}[theorem]{\sc Proposition}

\newtheorem{definition}[theorem]{\sc Definition}

\def\R{{\mathbf R}}

\def\D{{\mathcal D}}

\title{On the weak solutions of the McKendrick equation: Existence of demography cycles}

\author{Rui Dil\~ao\ \ and\ \  Abdelkader Lakmeche}

\begin{document}
\date{}

\maketitle

\begin{center}
Nonlinear Dynamics Group, Instituto Superior T\'ecnico \\
Av. Rovisco Pais, 1049-001 Lisbon, Portugal\\
{\tt rui@sd.ist.utl.pt; lakmeche@yahoo.fr}
\end{center}

\begin{abstract}
We develop the qualitative theory of the solutions of the McKendrick partial differential equation of population dynamics.
We calculate explicitly  the weak solutions  of the McKendrick equation and of the Lotka renewal integral equation with time and age dependent
birth rate. Mortality modulus  is considered age dependent. We show the existence of demography cycles. For a population with only one reproductive age class, independently of the stability of the weak solutions and after a transient time, the temporal evolution of the number of individuals of a population is always modulated by a time periodic function. The periodicity of the cycles  is equal to the age of the  reproductive age class, and 
a
population retains the memory from the initial data through the amplitude of oscillations.
For a population with a continuous distribution of reproductive age classes, the amplitude of oscillation is damped. The periodicity of the damped cycles is associated with the age of the first reproductive age class. Damping increases as the dispersion of the fertility function around the age class with maximal fertility increases. In general, the period of the
demography cycles  is associated with 
the time that a species takes to reach the reproductive maturity. 
\end{abstract}

\smallskip


\pagestyle{myheadings}
\thispagestyle{plain}

\section{Introduction}\label{s1}

The  McKendrick equation describes the time evolution of a population structured in age. The first time it appeared explicitly in the literature of population dynamics  was in 1926 in a paper by McKendrick \cite{b11}. The McKendrick equation is a first order hyperbolic partial differential equation, with time and age as independent variables, together with a boundary condition that takes into account the  births in a population. The existence of classical solutions of the McKendrick equation and their asymptotic time behaviour is well established, and there exists in the literature of population dynamics  a large number of surveys. See for example the books of  Cushing \cite{b2}, Webb \cite{b15}, Iannelli \cite{b6}, Keyfitz \cite{key}, Farkas \cite{farkas2}, Kot \cite{kot}, Charlesworth \cite{b1},   Metz  and  Diekmann \cite{b12}  and 
Chu \cite{chu}.    

Regardless the fact that the existence of classical solutions of the McKendrick equation is well established, there is a lack of specific examples and no explicit solutions of the McKendrick equation are known. This is due to the particular form of the boundary condition  which is difficult to handle analytically, \cite{kot} and \cite{farkas2}. 

The McKendrick modelling approach is an attempt to overcome the deficiencies shown by the Malthusian  or exponential growth law of population dynamics, introducing the
dependence on age into the mortality and fertility of a population. In its simpler form, the McKendrick model does not describe overcrowding effects, dependence on resources  or, in human populations, economic and intraspecific interactions. To include these effects, several other models have been introduced and analyzed from the mathematical and numerical point of view,  \cite{b15},   \cite{b12},  \cite{b3} and \cite{weiss}. 

In demography, in order to make predictions about population growth,  another approach is in general followed. After measuring birth and death rates  by age classes or cohorts, demographers  use the Leslie model \cite{b9}, a discrete analogue of the McKendrick equation, \cite{key} and \cite{b8}.

Our purpose here is to find the solutions of the McKendrick equation in the weak or distributional sense, to calculate exactly specific examples, and to derive some of their properties. To close the gap between theoretical and computational models, we compare the stability properties of the McKendrick and the Leslie discrete models, unifying both modelling approaches. 

From the point of view of the qualitative theory of partial differential equations,
we show that the solutions of the McKendrick equation have cycles, (demography or Easterlin cycles, \cite{easterlin} and \cite{chu}),
as observed in the growth of human \cite{b14} and bacterial populations \cite{b13}. 
In the Easterlin qualitative approach, \cite{easterlin} and \cite{key}, the period of the cycles
is estimated to be of the order of the age of two generations, which, for human population, is of the order of 50 years. Here, we show that the period of the demography cycles is associated with the age of the first reproductive age class, and, for human populations, is in the range 10-20 years.
The amplitudes of the cycles are damped and the damping is associated with the dispersion of the fertility  of a population around some maximal fertility age.  

In the next section, we review some of the facts about the 
McKendrick equation and we describe the methodology and organization of this  paper.

\section{Background and statement of results}\label{sback}
 
We denote by $n(a,t)$ the density of individuals of a population, where $a$ represents age and $t$ is time. 
Assuming that, within a population, death occurs with an age dependent mortality modulus $\mu (a)$,
we have, $\frac{dn(a,t)}{dt} =  - \mu (a)n(a,t)$. As aging is time dependent, $a\equiv a(t)$, and is measured within the same scale of time, $\frac{da}{dt}=1$,
the function $n(a,t)$ obeys the first order linear hyperbolic partial differential equation,
\begin{equation}
{\frac{\partial n(a,t)}{\partial t}} + {\frac{\partial n(a,t)}{\partial a}} 
=  - \mu (a) n(a,t)
\label{1.1}
\end{equation}
where $a\ge 0$ and $t\ge 0$. 
To describe births, an age specific fertility distribution function by age class $b(a)$ is introduced, and new-borns (individuals with age $a=0$) at time $t$ are calculated with the boundary condition,
\begin{equation}
n(0,t)=\int_{\alpha}^{\beta} b(a)n(a,t)da\, .    
\label{1.2}
\end{equation}
Supported by data from bacterial and human populations, \cite{b13} and \cite{b8}, $b(a)$ is a function with compact support in an
interval $[\alpha, \beta ]$, where $\alpha >0$ and $\beta <\infty $, and $\mu (a)$ is a non-negative function.  Equation (\ref{1.1}) together with the boundary condition (\ref{1.2}) is the McKendrick equation, \cite{b11}. Knowing the solutions of the McKendrick equation, the total population at time $t$ is given by,
\begin{equation}
N(t)=\int_0^{+\infty} n(a,t)da \, .
\label{1.3}
\end{equation}

The existence of solutions of the linear equation (\ref{1.1}) with boundary conditions (\ref{1.2}) has been implicitly proved by several authors and goes back to the work of McKendrick \cite{b11}, Lotka \cite{b10} and Feller \cite{b4}. More recently, Gurtin and MacCamy \cite{b5} proved the existence of solutions of (\ref{1.1}) for
a class of models where the mortality modulus depends on the total population, $\mu\equiv \mu (a,N(t))$.
However, due to the particular form of the boundary condition (\ref{1.2}), no explicit solutions of the McKendrick equation have been found. 

In 
the development of the theory of the McKendrick equation, the efforts have been on the derivation of existence of solutions, on  the conditions for asymptotic extinction and for positive equilibrium solutions, as well as, on the existence of stable age distributions in finite age intervals, \cite{b15} and  \cite{farkas2}.
From the point of view of demography, most of the conclusions derived from the McKendrick model are based 
on numerically constructed solutions from particular initial data.  

In the following,  we consider the more general case,
\begin{equation}
n(0,t)=\int_{\alpha}^{\beta} b(a,t)n(a,t)da   
\label{1.2a}
\end{equation}
where the fertility modulus $b(a,t)$ is  age and time dependent, and
for every $t\ge 0$, $b(a,t)$ has compact support in the interval $[\alpha,\beta]$, with $0<\alpha<\beta< \infty$. In order to simplify the calculations, we sometimes consider that the function $b(a,t)$ has a natural 
extension as a zero function to the half real line  $a\ge 0$.
This boundary condition is of special interest in demography and economy growth models, enabling the
analysis of the effect of fluctuations  of  the fertility modulus along time. 

To calculate 
the general solutions of the McKendrick equation (\ref{1.1}) subject to the boundary condition (\ref{1.2a}), we take the
initial data $n(a,0)=\phi (a)$, for $a>0$. In general,   $\phi(0)\not= \int_{\alpha}^{\beta} b(a,0)n(a,0)da$.  Under these conditions, as $t$ increases, the boundary condition (\ref{1.2})  or (\ref{1.2a})  introduces   discontinuities in the solutions of (\ref{1.1}). Here, we assume that 
$\phi (a)\in L^1_{loc}({\R}_+)$, implying that $n(a,t)$ is also locally integrable.
In this case, the total population at time $t=0$, $N(0)$, is  well defined only if $\phi (a)$ has compact support in ${\R}_+$. As we shall
see, this does not introduce any technical restriction because asymptotic solutions depend 
on $\phi (a)$ in an age interval of finite length.  
In this framework, the Cauchy problem for the McKendrick equation must be understood in the sense of distributions --- weak Cauchy problem.

To calculate explicitly the general solutions of the partial differential equation (\ref{1.1}) obeying
the boundary condition (\ref{1.2a}) and prescribed initial data,
we use the technique of characteristics, \cite{b7} and \cite{dilao2}. Writing equation (\ref{1.1}) in the form, ${{dn(a,t)} \over {dt}} =  - \mu (a) n(a,t)$,  the solutions of (\ref{1.1}) are also solutions of the system of ordinary differential equations,
\begin{equation}\left\{
\begin{array}{l}\displaystyle
   {{dn} \over {dt}} =  - \mu (a) n  \\ [6pt]
 \displaystyle  {{da} \over {dt}} = 1\, . \\
\end{array}\right.
\label{1.4}
\end{equation}
These two  equations have solutions,
\begin{equation}
\begin{array}{l}
n(a,t) = n(a_0 ,t_0 )\exp\left({ - \int_{t_0}^{t} \mu (s+a_0 - t_0 )ds}\right)  \\
 a - a_0  = t - t_0   \\
\end{array}
\label{1.5}
\end{equation}
where $a_0$ is the  age variable at time $t=t_0$. The second equation in (\ref{1.5}) is the equation of the characteristic curves of the partial differential equation (\ref{1.1}). Introducing the second equation in (\ref{1.5}) into the first one, we obtain the solution of the McKendrick equation for $t<a$,
\begin{eqnarray}
n(a,t) &=& \phi(a - t)\exp\left( { - \int_{0}^{t} \mu (s+a_0 )ds}\right)\nonumber\\
&=& \phi(a - t)\exp\left( { - \int_{a-t}^{a} \mu (s)ds}\right)\quad (t<a)
\label{1.6}
\end{eqnarray}
where $\phi(a-t)=n(a-t,0)$ is the initial age distribution of the population at the time $t_0=0$. For $t<a$, the solution (\ref{1.6}) does not depend on the boundary condition  (\ref{1.2a}), and if $\phi (a)\in L^1({\R}_+)$ or $\phi (a)\in L^1_{loc}({\R}_+)$, $n(a,t)$ is integrable or locally integrable in $a$, provided $\mu (a)\in L^1_{loc}({\R}_+)$ and $\mu (a)\ge 0$. 

For $t\ge a$, by (\ref{1.5}), we have,
\begin{equation}
n(a,t) = n(0,t - a)\exp\left({ - \int_{0}^{a} \mu (s)ds}\right):=n(0,t - a)\pi(a)
\label{1.7}
\end{equation}
and $\pi(a)$ can be understood as the probability of survival up to age $a$ of the individuals of the population.

In  mathematical demography, renewal theories of age-structured populations are in general used. Due to the simple structure of the characteristic curves of equation (\ref{1.1}),  Lotka \cite{b10} has shown that the density of newborns at time $t$,
$B(t):=n(0,t)$, obeys an integral equation. 
Introducing (\ref{1.6}) and (\ref{1.7}) into the boundary condition (\ref{1.2a}), and making the zero extension of $b(a,t)$ to the half real line $a\ge 0$, we obtain,
\begin{eqnarray} 
B(t)&=&\int_{0}^{t} b(a,t) \pi(a) B(t-a) da+ 
\int_{t}^{+\infty} b(a,t) \pi(a) \frac{\phi(a-t)}{\pi(a-t)} da \nonumber \\
&:=&\int_{0}^{t} {\bar b}(a,t) B(t-a) da+g(t)
\label{1.8}
\end{eqnarray}
which is the renewal integral equation of demography, first introduced by Lotka \cite{b10} and developed later by Feller \cite{b4}. 

If the solutions of integral equation (\ref{1.8}) are known, then, by (\ref{1.6}) and
(\ref{1.7}), 
the solutions of the McKendrick equation can be written as,
\begin{equation}
n(a,t)=\left\{ \begin{array}{l}
\frac{\pi(a)}{\pi(a-t)}\phi(a - t)\quad (t<a)\\[5pt]
\pi(a)B(t - a) \quad (t\ge a)\, .\\
\end{array}
\right.
\label{1.9}
\end{equation}
Therefore, once the solutions of the renewal integral equation (\ref{1.8}) are known, the solutions of the McKendrick equation (\ref{1.1}) with boundary condition (\ref{1.2a}) are readily derived. Inversely, if we solve the initial data problem for the McKendrick equation, the solution of the renewal equation is also easily derived. 

The time independent or equilibrium solutions of the McKendrick equation 
obey the ordinary differential equation ${{dn} \over {da}} =  - \mu (a) n $. Therefore, any equilibrium solution has the generic form, 
\[
n(a)=n_0\exp\left({ - \int_{0}^{a} \mu (s)ds}\right)
\]
where $n_0$ is a constant.
Multiplying the equilibrium solution by $b(a)$ and integrating in $a$, by
the boundary condition (\ref{1.2}), it follows that $\int_0^{ + \infty } b(a) e^{-\int_0^a\mu(s)ds}da=1$. Therefore, we define the Lotka growth rate as the number,
\begin{equation}
r=\int_0^{ + \infty } b(a) e^{-\int_0^a\mu(s)ds}da\, .
\label{Lot}
\end{equation}
It is a well known result that the properties of the asymptotic solutions of the McKendrick equation 
are determined by the Lotka growth number.

In the following two sections, we calculate explicitly the weak solutions of the McKendrick equation (\ref{1.1}), as well as the solutions of the renewal integral equation (\ref{1.8}).

In section \ref{s2}, we  assume that a population has only one fertile age class. This leads to the introduction of  a new boundary condition $n(0,t)=b_1(t)n (\alpha,t)$, where $b_1(t)$ is a positive function and $a=\alpha >0$ is the only reproductive age class. This corresponds to the choice $b(a,t)= b_1(t)\delta (a-\alpha)$, where $\delta (\cdot )$ is the Dirac delta function. In this case, the 
solutions of the McKendrick equation are easily derived and the renewal integral equation (\ref{1.8}) reduces to the functional equation,
\begin{equation}
B(t)=\left\{ \begin{array}{l}
b_1(t)\frac{\pi(\alpha )}{\pi(\alpha-t)}\phi(\alpha - t)\quad (t<\alpha)\\
b_1 (t)\pi(\alpha)B(t-\alpha) \quad (t\ge \alpha)\, .\\
\end{array}
\right.
\label{1.10}
\end{equation} 

The choice of the boundary condition $n(0,t)=b_1(t)n (\alpha ,t)$
has the advantage of describing the overall growth patterns of a population as if fecundity were concentrated in one reproductive age class. This enables to discuss effects associated with delays in reproduction and, as we shall see,  it is important to determine the   periods of the demography cycles. On the other hand, this boundary condition together with the linearity of
the McKendrick equation enables to unify the Leslie and McKendrick
approach, with considerable advantages for the qualitative understanding 
of growth effects in real populations.

One of the main results of section \ref{s2} is that the solution of the
McKendrick equation is the product of an  exponential function in time  by a periodic function with a period equal to the age of the only reproductive age class. The shape of the periodic modulation  depends on the initial age
distribution of the population, and the amplitude  depends on the mortality modulus.

In section \ref{s3}, we consider the general case where fertile age classes are distributed along an age interval, and we derive explicit formulas for the weak solutions of the McKendrick and the renewal equations. The asymptotic behaviour 
of the solutions is derived as a function of the Lotka growth rate.
Then,
we explicitly calculate the solutions of the renewal and of the McKendrick equation for the case where $b(a,t)$ is constant in the age interval $[\alpha,\beta]$.

  We show that 
the stability and instability of  the solutions of the McKendrick equation in the weak sense are determined by the Lotka growth rate.
 In the present theory of the McKendrick equation, the stability and instability of  the solutions 
are given implicitly by the real roots of the characteristic equation associated with the Laplace transform of (\ref{1.1}), \cite{b1}, \cite{b2} and \cite{b6}. An algorithmic procedure to determine the roots of the characteristic equation has been recently derived  by Farkas \cite{farkas}. However, it is not always possible to locate these roots, \cite{farkas}.  With the approach developed here, the stability and instability of solutions of the McKendrick equation are directly calculated  from  the fertility function $b(a)$ and the mortality modulus $\mu(a)$, parameters directly measured in demography, \cite{key} and \cite{b8}. 

The technique developed in section \ref{sML} enables to relate  and calibrate  the parameters  of both the McKendrick continuum model and the  Leslie discrete model, \cite{b9}, used in demography studies,
\cite{b8}. It is shown that when the number of age classes of the Leslie model goes to infinity, the inherent net reproductive number of a population converges to the Lotka growth rate.

In general, we conclude that  if the time intervals in the time series of a population equals the age of maximal fertility, this time series has an exponential or Malthusian growth. For smaller time intervals, the time series is modulated by a periodic function in time. 
This proves that, in a time scale of the order of 10-20 years, there exists cycling behaviour in the pattern of growth of a population. Approximating the age distribution of a population by the sum of Dirac delta functions concentrated at consecutive age classes, a modelling possibility from the computational point of view, the modulation of the pattern of growth of a populations becomes almost periodic in time (section \ref{sDP}).

At the end of each section, we discuss the qualitative aspects of the solutions we have analysed.  
The main conclusions of the paper for demography and population dynamics are discussed in the concluding section \ref{s4}. 

\section{Populations with one fertile age class}\label{s2}

We suppose that births occur at some fixed age $a=\alpha >0$,  and  we assume that the fertility function  is  $b(a,t)\equiv b_1(t) \delta (a-\alpha )$, where $\delta (\cdot )$  is the Dirac delta function, and $b_1(t)$ is a differentiable, positive and bounded function of  time. For this particular case, the
boundary condition (\ref{1.2a}) becomes,
\begin{equation}
n(0,t)=b_1(t)  n(\alpha ,t) 
\label{2.1}
\end{equation}
with $\alpha>0$. Note that, if $\alpha =0$, the Cauchy problem for the McKendrick equation is only determined
for $t\le a$. For $t>a$, by (\ref{2.1}), $n(0 ,t)$ becomes undetermined, 
as well as $n(a,t)$ with $t>a$.

\begin{figure}[htbp]
\centerline{\includegraphics[width=6.5 true cm]{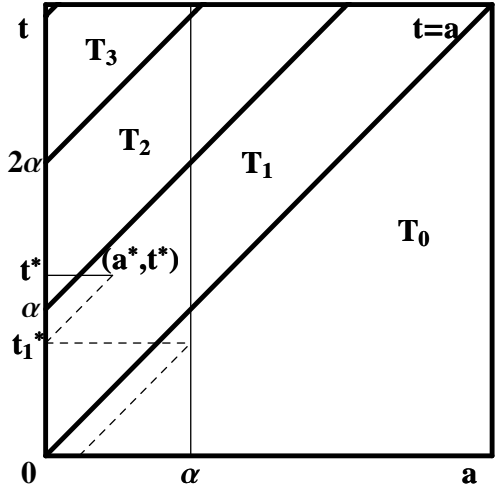}}
\caption{Characteristic curves $a - a_0  = t - t_0$ for the McKendrick equation (\ref{1.1}). The interior points of the sets ${\bar T}_m=\{(a,t):t<a+m\alpha, t\ge a+(m-1)\alpha ,a\ge 0, t\ge 0\}$, with $m\ge 1$, and ${\bar T_0}=\{(a,t):t<a,a\ge 0, t\ge 0\}$
are denoted by $T_m$, with $m\ge 0$. The vertical line $a=\alpha $ represents the age
of the unique reproductive age class of the population. The boundary condition
at time $t$ is calculated according to the value of $n(\alpha,t)$. At $t=0$, $n(a,0)=\phi (a)$. Given an arbitrary point $(a^*,t^*)$ in the domain of the partial differential equation (\ref{1.1}), $n(a^*,t^*)$  is  obtained following the solution $n(a,t)$ along the dotted line until $t=0$.}\label{fig1}
\end{figure}

To extend the solution (\ref{1.6}) of the McKendrick equation as a function of the initial data to all the domain of the independent variables $a$ and $t$, taking into account  the boundary condition (\ref{2.1}), we  first consider the case 
$t=a$.  By (\ref{1.7}), and for $t=a$,
$$
n(a,t) = n(0,0)\exp\left({ - \int_{0}^{a} \mu (s )ds}\right) \, .
$$
As newborns at time $t=0$ are calculated from the boundary condition (\ref{2.1}),  to make this solution dependent on the initial data,  we must have,
\begin{equation}
n(a,t) =  b_1(0)\phi(\alpha)\exp\left({ - \int_{0}^{a} \mu (s)ds}\right)\, ,\quad   \hbox{if} \quad  t = a 
\label{2.2}
\end{equation}
where $n(\alpha ,0)=\phi(\alpha)$, see Figure \ref{fig1}.

We now introduce  the sets ${\bar T}_m=\{(a,t):t<a+m\alpha, t\ge a+(m-1)\alpha ,a\ge 0, t\ge 0\}$, where $m$ is a positive integer, and  ${\bar T_0}=\{(a,t):t<a,a\ge 0, t\ge 0\}$, Figure \ref{fig1}. We denote 
by $T_m$ the interior  of the sets ${\bar T}_m$. So, given the point
with coordinates $(a,t)$, we have, $(a,t)\in {\bar T}_m$, with $m\ge 1$, if and only if, $t\ge a$ and $[1+(t-a)/\alpha]=m$, where we have used the notation $[x]$  for the integer part of $x$. A point $(a,t)\in {\bar T}_0$, if and only if, $t<a$. Note that, the domain of the solution (\ref{1.6}) is the interior of the region labelled $T_0$ in Figure \ref{fig1}.

For $t> a$ and $t< a+\alpha $,  we take the point $(a^*,t^*)$  on the line $t=t^*$.  Therefore, $t^*> a^*$ and $t^*< a^*+\alpha $. This point is in the region labelled $T_1$ in Figure \ref{fig1}. By (\ref{1.5}), the characteristic line that passes by $(a^*,t^*)$   crosses the line $a=0$ at some time $t=t_1^*$, and, 
$$
n(a^*,t^*) = n(0 ,t_1^*)\exp\left( { - \int_{t_1^*}^{t^*} \mu (s-t_1^* )ds} \right)
$$
where $t_1^* = t^* - a^*$. Imposing the boundary condition (\ref{2.1}) on this solution,  we obtain,  $n(a^*,t^*) = b_1(t_1^*)n(\alpha ,t_1^*)\exp\left({ - \int_{t_1^*}^{t^*} \mu (s-t_1^* )ds}\right)$. 
By hypothesis, as $t_1^*=(t^* - a^*)< \alpha$, we are in the conditions of solution (\ref{1.6})
for $t=t_1^*$ and $a=\alpha$, 
and we have,
\begin{eqnarray*}
n(a^*,t^*) &=& b_1(t^* - a^*) n(\alpha ,t^* - a^*)\exp\left({ - \int_{t_1^*}^{t^*} \mu (s-t_1^* )ds}\right)\\ 
&=& b_1(t^* - a^*) \phi(\alpha  + a^* - t^*)  \\
&\times&\exp\left({ -  \int_{0}^{t^* - a^*} \mu (s+a_0)ds-\int_{t^* - a^*}^{t^*} \mu (s+a^* - t^* )ds }\right)\\ 
&=& b_1(t^* - a^*) \phi(\alpha  + a^* - t^*) \\
&\times&\exp\left({ -  \int_{0}^{t^* - a^*} \mu (s+\alpha +a^*-t^*)ds}\right) \\
&\times&\exp\left({-\int_{t^* - a^*}^{t^*} \mu (s+a^* - t^* )ds }\right)\, .\\ 
\end{eqnarray*}
Therefore,  we have shown that, for $t \ge a$ and  $t< a+\alpha $,
\begin{equation}
\begin{array}{ll}
n(a,t) =& b_1(t-a) \phi(\alpha  + a  - t )\\[5pt]
&\times \exp\left(\displaystyle
{ -  \int_{0}^{t - a} \mu (s+\alpha+a-t)ds-\int_{0}^{a} \mu (s)ds }\right)  .\\
\end{array}
\label{2.3}
\end{equation}
Note that, for $t\le a$, the solution $n(a,t) $ given by (\ref{1.6}), (\ref{2.2}) and (\ref{2.3}) is in general discontinuous when the line $t=a$ is crossed transversally.

We now proceed by induction.
Suppose that, up to some integer $k\ge 1$, the solutions  of the McKendrick equation (\ref{1.1}) with the boundary condition (\ref{2.1}) can be written in the form,
\begin{equation}
\begin{array}{ll}\displaystyle
n(a,t) =& \phi(m\alpha  + a  - t )\displaystyle \prod_{i=1}^{m}b_1(t-a- (i-1)\alpha)\\[7pt] 
&\times \exp\left(\displaystyle 
{ -  \int_{0}^{t - a-(m-1)\alpha} \mu (s+m\alpha+a-t)ds}\right) \\ [7pt] 
&\times \exp\left(\displaystyle
{-(m-1)\int_{0}^{\alpha} \mu (s)ds-\int_{0}^{a} \mu (s)ds }\right)   
\end{array}
\label{2.4}
\end{equation}
where $m=[(t-a)/\alpha +1]$, $(a,t)\in {\bar T}_m$, $m\le k$ and $m\ge 1$. As we have shown, (\ref{2.4}) is true for  $m=1$.
Suppose now that $(a^*,t^*)\in T_{k+1}$.
Then, by (\ref{1.5}), the characteristic curve that passes by $(a^*,t^*)$ crosses the line $a=0$ at some  time $t_1^*=t^*-a^*$, and,
\begin{eqnarray*} 
n(a^*,t^*)=n(0,t_1^*)\exp\left({ - \int_{t_1^*}^{t^*} \mu (s-t_1^* )ds}\right)\, .
\end{eqnarray*}
Imposing the boundary condition (\ref{2.1}) on this solution, we obtain,
\begin{eqnarray*} 
n(a^*,t^*)=b_1(t_1^*)n(\alpha,t_1^*)\exp\left({ - \int_{t_1^*}^{t^*} \mu (s-t_1^* )ds}\right)\, .
\end{eqnarray*}
As $[(t^*-a^*-\alpha)/\alpha +1]=[(t^*-a^*)/\alpha]=k$, we have, $(\alpha,t_1^*)\in {\bar T}_{k}$, and by (\ref{2.4}), 
\begin{eqnarray*}
n(a^*,t^*)&=&b_1(t_1^*)n(\alpha,t_1^*)\exp\left({ - \int_{t_1^*}^{t^*} \mu (s-t_1^* )ds}\right)\\ &=&
b_1(t^*-a^*)n(\alpha,t^*-a^*)\exp\left({ - \int_{0}^{a^*} \mu (s)ds}\right)\\
&=&\left(\prod_{i=0}^{k}b_1(t-a- i\alpha) \right)
 \phi((k+1)\alpha  + a^*  - t^* )\exp\left({ - \Xi}\right)\\
\end{eqnarray*}
where,
\begin{eqnarray*}
\Xi &=& \int_{0}^{t^* - a^*-k\alpha} \mu (s+(k+1)\alpha  + a^*  - t^* )ds \\
&+& k\int_{0}^{\alpha} \mu (s)ds+\int_{0}^{a^*} \mu (s)ds \, .\\
\end{eqnarray*}
Hence, (\ref{2.4}) remains true for $m=k+1$, and we have:

\begin{proposition}\label{P2.1}
Let  $n(a,0)=\phi(a )\in C^{1}({\R}_+)$  be an initial condition for the McKendrick partial differential equation (\ref{1.1}), with $a\ge 0$, $t\ge 0$, and boundary condition (\ref{2.1}). Assume that $\mu (a)\in C^{0}({\R}_+)$ is a non negative function and $b_1 (t)\in C^{1}({\R}_+)$ is positive. 
Then, in the interior of the sets ${\bar T}_m$, with $m\ge 0$, the solution of the McKendrick equation (\ref{1.1})  is differentiable in $a$ and $t$ and is given by:

\noindent {a)} If,  $(a,t)\in  T_0$,
$$
n(a,t) =  \phi(a - t)\exp\left({ - \int_{0}^{t} \mu (s+a_0)ds}\right)
$$
where $a_0=a-t$.

\noindent {b)} If, $(a,t)\in  T_m$ and $m\ge 1$,
\begin{eqnarray*}
n(a,t) &=& \prod_{i=1}^{m}b_1(t-a- (i-1)\alpha)  \\
& &\times \phi(m\alpha  + a - t)\exp\left({- \int_{0}^{t - a-(m-1)\alpha} \mu (s+a_0)ds}\right)  \\
& &\times \exp\left({-
(m-1)\int_{0}^{\alpha} \mu (s)ds-\int_{0}^{a} \mu (s)ds) }\right)
\end{eqnarray*}
where $a_0=m\alpha  + a - t$, $m=\left[ {(t - a)/\alpha  + 1} \right]$, $[x]$ stands for the integer part of $x$, and  $\alpha >0$.
\end{proposition}

\begin{proof}
As $\phi(a )\in C^{1}({\R}_+)$, $\mu (a)\in C^{0}({\R}_+)$ and
 $b_1 (t)\in C^{1}({\R}_+)$, it is straightforwardly checked by differentiation that, in the interior of the sets ${\bar T}_m$, with $m\ge 0$, the solution obtained by the method of characteristics obeys the McKendrick equation (\ref{1.1}).
\end{proof}

In the construction preceding Proposition \ref{P2.1},
we have shown that the solutions of the McKendrick equation  hold formally at the boundary of the sets ${\bar T}_m$, $m\ge 0$, where $n(a,t)$ is discontinuous because,  in general, $\phi(0)\not= b_1(0)\phi(\alpha )$.
To extend the solution of the equation (\ref{1.1}) as stated in  Proposition \ref{P2.1} to all the domain of the independent variables,
we introduce the concept of weak solution in the sense of distributions, \cite{b7}. We consider the
space of test functions that is, the space of functions of compact support in 
${\R}^2_+$ with derivatives of all the orders. 
Let ${\D}({\R}^2_+)$ be the space of test functions. If $f(x)$ is a locally integrable function, then $f[\psi]=\int \psi(x)f(x) dx$ is a continuous functional in the sense of the distribution in ${\D}({\R}^2_+)$, \cite{b7}. Taking  equation (\ref{1.1}) and making the inner product with a function $\psi(a,t) \in {\D}({\R}^2_+)$, as $\psi $
has compact support in ${\R}^2_+$, and after integrating by parts, we  naturally arrive
at the following definition:

\begin{definition}\label{D2.2}
A locally integrable function $n(a,t)\, (\in L^1_{loc} ({\R}^2_+))$ is a weak solution in the sense of the distributions of the  McKendrick partial differential equation (\ref{1.1}),  if,  
$$
\int \int_{{\R}^2_+}  \left( {\partial \over \partial t}\psi (
a,t) +{\partial \over \partial a}\psi ( a,t)-\mu(a) \psi ( a,t)\right) n(a,t) da dt=0
$$
for any $\psi(a,t) \in {\D}({\R}^2_+)$.
\end{definition}

In the conditions of Definition \ref{D2.2}, we have:

\begin{theorem}\label{T2.3} 
Let  $n(a,0)=\phi(a )\, (\in L^1_{loc} ({\R}_+))$ be a locally integrable initial condition for the McKendrick partial differential equation (\ref{1.1}), with $a\ge 0$, $t\ge 0$, and boundary condition (\ref{2.1}). Assume that $\mu (a)\in L^1_{loc} ({\R}_+)\cap C^{0}({\R}_+)$ is a non negative function and $b_1 (t)\in C^{1}({\R}_+)$ is positive. Then, the weak  solutions of the McKendrick equation (\ref{1.1})  are:

\noindent {a)} If,  $(a,t)\in  {\bar T}_0$,
$$
n(a,t) =  \phi(a - t)\exp\left({ - \int_{0}^{t} \mu (s+a_0 )ds}\right)
$$
where $a_0=a-t$.

\noindent {b)} If, $(a,t)\in  {\bar T}_m$ and $m\ge 1$,
\begin{eqnarray*}
n(a,t) &=& \prod_{i=1}^{m}b_1(t-a- (i-1)\alpha )  \\
& & \times \phi(m\alpha  + a - t)\exp\left({ -  \int_{0}^{t - a-(m-1)\alpha} \mu (s+a_0)ds}\right) \\
& &\times \exp\left({ -
(m-1)\int_{0}^{\alpha} \mu (s)ds-\int_{0}^{a} \mu (s)ds }\right)
\end{eqnarray*}
where  $a_0=m\alpha  + a - t$, $\alpha >0$,  $m=\left[ {(t - a)/\alpha  + 1} \right]$, and $[x]$ stands for the integer part of $x$.
\end{theorem}

\begin{proof}
We have shown previously that the solutions of the McKendrick equation as in Proposition \ref{P2.1} hold at the boundary of the sets ${\bar T}_m$, with $m\ge 1$. However, at these boundary points the solutions in Proposition \ref{P2.1} are not differentiable and  must be understood   as
weak solution in the sense of distributions.
So, for $n(a,t)$ as in Proposition \ref{P2.1} to be a solution of the McKendrick equation
in all the domain of the independent variables, it must satisfies the  conditions in
Definition \ref{D2.2}  for any $\psi(a,t) \in {\D}( {\R}_+^{2})$. As by hypothesis $\phi(a)$ and $\mu (a)$ are locally integrable, we are in the conditions of Definition \ref{D2.2}, and we can verify if $n(a,t)$ given by Proposition \ref{P2.1} can be extended as a weak solution in all the domain of $a$ and $t$. For that, we write the solution in a) in Proposition \ref{P2.1} 
in the form, 
$n(a,t)=\Phi (a-t)  e^{-\Upsilon (a-t,t)}$, where $\Upsilon (a-t,t)=\Upsilon (a_0,t)$ is a
generic function reflecting the functional dependency of the exponential term, and we calculate the integral,
$$
I=\int \int_{{\R}^2_+}  \left( {\partial \psi \over \partial t} + {\partial \psi \over \partial a}  - \mu(a) \psi\right) \Phi (a-t) e^{-\Upsilon (a-t,t)} da dt 
$$
where $  \Phi(y)=0$, for $y<0$, $\Phi(y)=\phi(y)$, for $y\ge0$, and $\Phi(y)\in \ L^1_{loc} ({\R})$.
Introducing the new coordinates $y=a-t$ and $x=t$, we have, 
$$
I=\int \int_{{\R}\times {\R}_+} \left( {\partial \psi \over \partial x}   - \mu(x+y) \psi\right) \Phi (y) e^{-\Upsilon (y,x)} dx dy\, .
$$
As, $\Upsilon (a-t,t)=\Upsilon (y,x)=\int_0^x \mu(s+y)\, ds$, and ${\partial \Upsilon \over \partial x} =\mu(x+y)=\mu (a)$, almost everywhere,  the above integral evaluates to,
$$
I=\int \int_{{\R}\times {\R}_+}{\partial   \over \partial x}\left(\psi e^{-\Upsilon (y,x)}\right)\Phi (y)dxdy
=\int_{{\R}}\left[\psi e^{-\Upsilon (y,x)}\right]_0^{\infty}\Phi (y)dy=0\, .
$$
Therefore, the solution a) is a weak solution of the McKendrick equation. For the case b),
we introduce the new coordinates $y=t-a$ and $x=t$. Writing the solution in b) 
in the form, 
$n(a,t)=\Phi (t-a)  e^{-\Upsilon (t-a,t)}$, and as $m\equiv m(y)$, it follows that ${\partial \Upsilon \over \partial x} =\mu(x-y)=\mu (a)$. By an analogous calculation, we have $I=0$, and 
the solution b) is also a weak solution of the McKendrick equation.
\end{proof}

For the particular case $a=0$, Theorem \ref{T2.3} gives the explicit solution of the renewal 
functional equation (\ref{1.10}).

\begin{corollary}\label{C2.4} 
Let  $n(a,0)=\phi(a )\, (\in L^1_{loc} ({\R}_+))$ be a locally integrable initial condition for the McKendrick partial differential equation (\ref{1.1}), with $a\ge 0$, $t\ge 0$. Assume that $\mu (a)\in L^1_{loc} ({\R}_+)\cap C^{0}({\R}_+)$ is a non negative function and $b_1 (t)\in C^{1}({\R}_+)$ is positive. Then the solution of the renewal functional equation (\ref{1.10}) is,
\begin{eqnarray*}
B(t) &=& \phi(m\alpha - t) \prod_{i=1}^{m}b_1(t- (i-1)\alpha)  \\
& &\times  \exp\left({ -  \int_{0}^{t -(m-1)\alpha} \mu (s+m\alpha  - t)ds}\right) \\
& &\times \exp\left({-(m-1)\int_{0}^{\alpha} \mu (s)ds }\right)
\end{eqnarray*}
where $\alpha >0$,  $m=\left[ {t/\alpha  + 1} \right]$, and $[x]$ stands for the integer part of $x$.
\end{corollary}

In the particular case where $b_1(t)$ is a  constant,   $b_1(t)\equiv b_1$, we have, in Theorem \ref{T2.3} and Corollary \ref{C2.4},  $b_1(t-a) \ldots b_1(t-a- (m-1)\alpha)=b_1^m$.
If $\mu(a)\equiv \mu >0$ and $b_1(t)\equiv b_1>0$ are constant functions, the solution of the McKendrick equation is,
\begin{eqnarray*}
n(a,t)=\left\{ \begin{array}{l}
\phi(a - t)e^{ - \mu t}\quad (t<a) \\
b_1^{m} \phi(m\alpha  + a - t)e^{ - \mu t}\quad (t\ge a)\\
\end{array}\right.
\end{eqnarray*}
where $m=\left[ {(t-a)/\alpha  + 1} \right]$.
For the same case, the solution of the renewal equation (\ref{1.10}) is,
$$
B(t) = b_1^{s} \phi(s\alpha - t)e^{ - \mu t}
$$
where $s=\left[ {t/\alpha  + 1} \right]$.

To analyze the  asymptotic behaviour in time of the solution of the McKendrick equation with time independent fertility function, we   define the function $\varepsilon (a,t)$ by, 
\begin{equation}
\varepsilon (a,t)=(t - a)/\alpha  + 1 - m=(t - a)/\alpha  + 1-\left[ {(t - a)/\alpha } +1\right] 
\label{2.5}
\end{equation}
where  $\left[ {(t - a)/\alpha } +1\right] = m$, and $m\ge 1$  is an integer.
For fixed $a$ and with $t-a\ge 0$, the function $\varepsilon (a,t)$ takes values in $[0,1)$, is piecewise linear and is time periodic with period $\tau =\alpha$. Then, we have:

\begin{theorem}\label{T2.5} 
Suppose that $\phi(a)$ is positive and bounded in the interval $(0,\alpha ]$ and the fertility function $b_1$ is a positive constant.
Then, in the conditions of Theorem \ref{T2.3}, we have:

\noindent {a)} If, $\ln b_1 =\int_0^{\alpha}\mu(s)\, ds$, then, for fixed $a$ and $t\ge a$,
$n(a,t)$ is time periodic with period $\tau =\alpha$.

\noindent {b)} If, $\ln b_1 >\int_0^{\alpha}\mu(s)\, ds$, then, for fixed $a$ and  $t\ge a$, $n(a,t)\rightarrow \infty $, as  $t\rightarrow \infty $.

\noindent {c)} If, $\ln b_1 <\int_0^{\alpha}\mu(s)\, ds$, then, for fixed $a$ and  $t\ge a$, $n(a,t)\rightarrow 0 $, as  $t\rightarrow \infty $.
Moreover, for fixed $a$, the asymptotic behaviour in time of $n(a,t)$ depends on the initial condition $\phi(a)$ with $a$ in the interval $(0,\alpha]$. 
\end{theorem}

\begin{proof}
By (\ref{2.5}), with $a_0=m\alpha  + a - t$,
\begin{eqnarray*}
&b_1^{m}&\exp\left({ -  \int_{0}^{t - a-(m-1)\alpha} \mu (s+a_0)ds-
(m-1)\int_{0}^{\alpha} \mu (s)ds}\right)  \\
&\times&\exp\left({-\int_{0}^{a} \mu (s)ds }\right)\\
&=&\exp\left({m(\ln b_1-\int_0^{\alpha} \mu (s)\, ds)} \right)
\exp\left({-\int_{0}^{a} \mu (s)ds }\right)\\
&\times&\exp\left({ -  \int_{0}^{\alpha\varepsilon(a,t)} \mu (s+\alpha-\alpha\varepsilon(a,t))ds
+\int_{0}^{\alpha} \mu (s)ds}\right)  \\
&:=&\exp\left({m(\ln b_1-\int_0^{\alpha} \mu (s)\, ds)}\right) 
\exp\left(-{\int_{\alpha}^a \mu (s)\, ds}\right) \chi(a,t)
\end{eqnarray*}         
where,
\begin{eqnarray}
\chi(a,t)&=&
\exp\left({ -  \int_{0}^{\alpha\varepsilon(a,t)} \mu (s+\alpha-\alpha\varepsilon(a,t))ds}\right)
\nonumber \\
&=&
\exp\left({ -  \int_{\alpha-\alpha\varepsilon(a,t)}^{\alpha} \mu (s)ds }\right)\, .
\label{2.8}
\end{eqnarray}
Due to the periodicity of $\varepsilon(a,t)$ in $t$, for fixed $a$ and $t\ge a$, $\chi(a,t)$ is also periodic in $t$, and we can write the solution b) of Theorem \ref{T2.3} in the form,
\begin{eqnarray} 
n(a,t) &=&  \phi(\alpha - \alpha \varepsilon (a,t))
\chi(a,t)   \nonumber \\
&\times & \exp\left({m(\ln b_1-\int_0^{\alpha} \mu (s)\, ds)}\right) 
\exp\left(-{\int_{\alpha}^a \mu (s)\, ds}\right)  \,.
\label{2.6}
\end{eqnarray}
Therefore, if $\ln b_1 -\int_0^{\alpha}\mu(s)\, ds= 0$ and $t\ge a$,
the solution of the McKendrick partial differential equation  becomes oscillatory in time. If, $\ln b_1 >\int_0^{\alpha}\mu(s)\, ds$,
then, in the limit $t\to \infty$, $m\to \infty$, and the population density goes to infinity. If, $\ln b_1 <\int_0^{\alpha}\mu(s)\, ds$, then, in the limit $t\to \infty$, the population density goes to zero. By (\ref{2.6}),  the asymptotic distribution of a population depends on $\phi (a)$ with $a \in (0,\alpha ]$.
\end{proof}

If we fix an arbitrary large value of
$a$, and if $\ln b_1 >\int_0^{\alpha}\mu(s)\, ds$, then $n(a,t)\rightarrow \infty $, as  $t\rightarrow \infty $, implying that asymptotically there is not a limit in life expectancy.

Due to the periodicity in time of $\varepsilon(a,t)$, the functions
$\chi(a,t)$ and $\phi(\alpha - \alpha \varepsilon (a,t))$ are time periodic with period $\tau =\alpha$, the age of the only reproductive age class.

Defining the Lotka growth rate of a population as
$r=n(a,t+\alpha )/n(a,t)$, by Theorem \ref{T2.3} and as $\left[ {(t+\alpha - a)/\alpha  + 1} \right]=m+1$, the Lotka growth rate associated with the McKendrick equation is constant and is given by,
\begin{equation}
r={n(a,t+\alpha )\over n(a,t)}=\exp\left({\ln b_1-\int_0^{\alpha} \mu (s)\, ds}\right)=
b_1\exp\left({-\int_0^{\alpha} \mu (s)\, ds}\right)
\label{2.9}
\end{equation}
which coincides with (\ref{Lot}) for the choice $b(a)=b_1\delta(a-\alpha)$.

The Lotka growth rate (\ref{2.9}) has a simple interpretation. By Theorem \ref{T2.3} and (\ref{2.9}),  for $t\ge a$, we have $n(a,t+s \alpha)= n(a,t)  r^{s}$, where $s$ is an integer. With $t=\alpha$ and 
$\alpha+s \alpha = \tau$,  we obtain, $n(a,\tau)=n(a,\alpha) r^{(\tau-\alpha)/\alpha}$. Defining the Malthusian density growth function as $n_M(a,t)=n(a,\alpha) r^{(t-\alpha)/\alpha}$, and integrating $n_M(a,t)$ in $a$,
the total population varies in time according to,
\begin{equation}
N_M(t)=N(\alpha) r^{(t-\alpha)/\alpha}
\label{2.10}
\end{equation}
which is the Malthusian growth function associated to the McKendrick equation. Therefore, within an observation time step equal to the age of the only reproductive age class,
the solutions of the McKendrick equation grow exponentially in time, with the Lotka growth rate (\ref{2.9}). 

The qualitative difference between the asymptotic behaviour of the solutions of the McKendrick equation and the simple Malthusian exponential growth without age structure  is associated with  
the existence of a periodic modulation  in the growth of populations. This periodic modulation with period $\alpha $ corresponds to the demography cycles observed in real populations, \cite{b13} and   
\cite{b14}. 

The stability condition  for the persistence of non-zero solutions as stated in Theorem \ref{T2.5} is determined by the growth rate $r$: stability or periodicity of solutions if $r=1$; exponential growth if $r>1$, and population extinction if $r<1$, provided $\phi (a)$ is not identically zero in $(0,\alpha]$. 

\begin{theorem}\label{T2.6} 
If $\phi(a)$ is bounded  in the interval $(0,\alpha ]$ and the fertility function $b_1$ is a positive constant, then, 
in the conditions of Theorem \ref{T2.3}, 
for fixed $a$,  and $t\ge a$,
$$
{n(a,t)\over r^m}=\chi(a,t)\phi(\alpha - \alpha \varepsilon (a,t))\exp\left({-\int_{\alpha}^{a} \mu (s)\, ds}\right)
$$
is a time periodic function with period $\alpha$, and $m=\left[ {(t - a)/\alpha  + 1} \right]$.  
\end{theorem}
  
\begin{proof}
 The theorem follows by (\ref{2.5}), (\ref{2.8}), (\ref{2.6}) and (\ref{2.9}). 
\end{proof}

By the above Theorem, the population age density and in the total population  has a periodic modulation in time.   These oscillations occur around the Malthusian growth curve (\ref{2.10}).

\begin{figure}[htbp]
\centerline{\includegraphics[width=13.0 true cm]{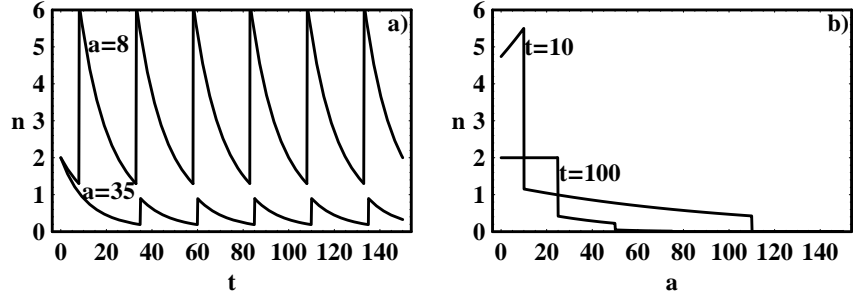}}
\caption{a) Time evolution of the solution of the McKendrick equation (\ref{1.1}) for age classes $a=8$  and $a=35$, in a population with one reproductive age class $\alpha =25$. The mortality modulus is $\mu (a)=0.05+0.001 a$ and $b_1=4.77$. b) Distribution of the density of individuals as a function of age for $t = 10$ and $t = 100$. In both cases, we have the stability condition 
$b_1 = e^{\mu \alpha}$, implying that the Lotka growth rate is $r=1$. All the  solutions have been calculated from Theorem \ref{T2.3},  with the initial data condition $\phi(a) = 2$,
for $a\le 100$, and $\phi(a) = 0$, for $a> 100$.}
\label{fig2}
\end{figure}

The amplitude of oscillations of the function ${n(a,t)/ r^m}$ can be easily determined. 
From Theorem \ref{T2.6}  the amplitude of oscillations is,
\begin{equation}
\begin{array}{lcl}
A&=&\max_{a\in(0,\alpha]}\chi(a,t)\phi(\alpha - \alpha \varepsilon (a,t))
\exp\left({-\int_{\alpha}^{a} \mu (s)\, ds}\right) \\[5pt]
&-&\min_{a\in(0,\alpha]}\chi(a,t)\phi(\alpha - \alpha \varepsilon (a,t))\exp\left({-\int_{\alpha}^{a} \mu (s)\, ds}\right)\, .
\end{array}
\label{2.11}
\end{equation}
Assuming that the initial distribution of the population is constant in the interval $(0,\alpha ]$, and as $\varepsilon (a,t)$
takes values in $[0,1)$, $n(a,t)/r^m$ varies in the interval, 
\[
\left[ \phi e^{-\int_{0}^{\alpha} \mu (s)\, ds}e^{-\int_{\alpha}^{a} \mu (s)\, ds},\phi e^{-\int_{\alpha}^{a} \mu (s)\, ds}\right]\, .
\]
Then, for a uniform initial age density of individuals, the periodic modulation has the age dependent amplitude, 
\begin{equation}
A(a)=\phi \exp\left({\int_{a}^{\alpha} \mu (s)\, ds}\right)
\left( 1-\exp\left({-\int_{0}^{\alpha} \mu (s)\, ds}\right) \right)\, .
\label{2.12}
\end{equation}
In the particular case where $\mu (a)\equiv \mu$,  the amplitude simplifies further and is given by $A(a)=\phi e^{\mu (\alpha -a)}(1-e^{-\mu \alpha })$. Therefore, the newborns age-class has the maximum amplitude of oscillations,
$A(0)=\phi e^{\mu \alpha }(1-e^{-\mu \alpha })$, and the amplitude of oscillations increases (resp. decreases) when $\mu $ increases (resp. decreases).  

Another consequence of Theorem \ref{T2.6} is that, in the limit $t\to \infty$, the
population retains the memory from the initial data through the amplitude of oscillations of the growth cycles, (see the discussion in
Iannelli \cite{b6}, pp. 37).

In Figure \ref{fig2}, we depict the time and age evolution of the density $n(a,t)$ of a population from a uniform initial age distribution with a maximal age class, and Lotka growth rate $r=1$. We have chosen the age-dependent mortality modulus $\mu (a)=\mu_0+\mu_1 a$, and as initial condition the density function with compact support, $\phi(a) = 2$, for $a\le 100$, and  $\phi(a) = 0$, for $a> 100$. By (\ref{1.3}), this corresponds to an initial population $N(0)=200$. 
The birth constant $b_1$ has been chosen in such a way that $r=1$. After the transient time $\alpha$, and
for fixed age, the population density becomes periodic in time with period $\tau =\alpha $. In Figure \ref{fig2}a, the decrease of the amplitude of oscillations with increasing $\mu (a)$ is in
agreement with (\ref{2.12}).

\begin{figure}[htbp]
\centerline{\includegraphics[width=13 true cm]{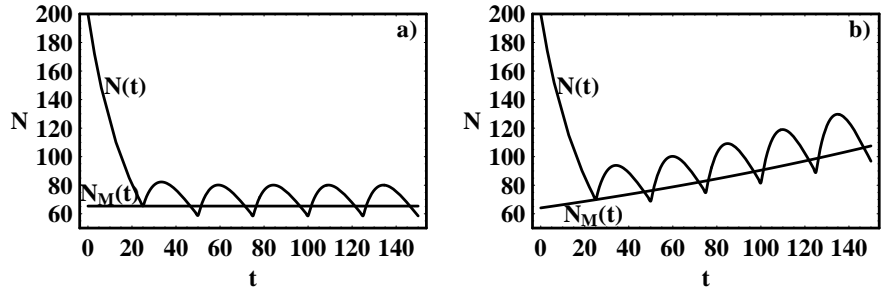}}
\caption{Total population number $N$ as a function of time calculated from Theorem \ref{T2.3} and (\ref{1.3}).  The age of the only reproductive age class is $\alpha =25$ and the initial conditions
are as in Figure 2. In a),  $\ln b_1=\int_0^{\alpha} \mu(s) ds=4.77$, where $\mu (a)=0.05+0.001 a$, $b_1=4.77$ and the Lotka growth rate or stability condition is $r=1$. In b), $b_1=5.2$, the Lotka growth rate is $r=1.09$ and 
the number of individuals of the population goes to infinity. We also depict the Malthusian growth function (\ref{2.10}), $N_M(t)$, measured in the time scale of the unique reproductive age class. In all the cases and after a transient time, the growth curve of the population is modulated by a periodic function with period $\tau =\alpha$.}
\label{fig3}
\end{figure}

In Figure \ref{fig3}, we show the total population $N$ as a function of time calculated from (\ref{1.3}) and Theorem \ref{T2.3}. In Figure \ref{fig3}a), the Lotka growth rate is $r=1$.
In Figure \ref{fig3}b), $b_1>\exp(\int_0^{\alpha} \mu(s) ds)$, and the Lotka growth rate is $r=1.09$. In both cases, we compare the solution $N(t)$ calculated from Theorem \ref{T2.3} and the Malthusian growth curve (\ref{2.10}). 

If the fertility function $b_1$ as considered in Theorem \ref{T2.5} depends on time, in general, the stability can not be decided in finite time. This follows by a
similar analysis to the  one in the proof of Theorem \ref{T2.5}.

The main conclusion about the solutions of the McKendrick equations for the 
boundary condition (\ref{2.1}) with $b_1(t)\equiv b_1$ is that the pattern of growth of a population
is  modulated by a periodic function with a period equal to the age of the only reproductive age class. The shape of the periodic modulation depends on the initial age
distribution of the population, and the amplitude depends on the mortality modulus. 
In time steps of the order of the age of the first reproductive class, the growth is Malthusian. Pure Malthusian growth is obtained if
the mortality modulus approaches zero
and the population has an uniform initial age distribution.

\section{Populations with several fertile age-classes}\label{s3}

We  consider now that the fertile ages of the individuals of a population are distributed in some age interval $[\alpha ,\beta ]$, with $0<\alpha  < \beta <\infty $.  The  boundary condition  is now,
\begin{equation} 
n\left( 0,t\right) =\int_{\alpha }^{\beta }b\left( a,t\right) n\left( a,t\right) da\, .
\label{3.1}
\end{equation}
The constants $\alpha $ and $\beta $ represent the ages of the first and the last reproductive age classes, respectively.

 As in (\ref{1.6}), if, $t<a$, the solution of the McKendrick equation $(\ref{1.1})$ is given by,	
\begin{equation} 
n\left( a,t\right) =\phi(a - t)e^{ - \int_{0}^{t} \mu (s+a-t )ds} 
\label{3.2}
\end{equation}
where $\phi (a)=n( a,0)$ is the initial age distribution of the population. 

The general solution of the McKendrick equation in all the domain of the independent variables
is obtained in the following way:

\begin{theorem}\label{T3.1} 
Let  $n(a,0)=\phi(a )\, (\in L^1_{loc} ({\R}_+))$ be a locally integrable initial condition for the McKendrick partial differential equation (\ref{1.1}), with $a\ge 0$, $t\ge 0$, and boundary condition (\ref{3.1}). Assume that $\mu (a)\in L^1_{loc} ({\R}_+)\cap C^{0}({\R}_+)$ is a non negative function, and that, for every $t\ge 0$, $b(a,t)$ is locally integrable and differentiable. Then, 
in the strip $S=\{(a,t):a\ge 0, 0\le t\le \alpha \}$, the weak  solution of the McKendrick equation (\ref{1.1}) 
with initial data $\phi(a)$ and boundary condition (\ref{3.1}) is:
\begin{equation}
n_S(a,t) =\left\{ \begin{array}{l} 
\phi(a - t)e^{ - \int_{a-t}^{a} \mu (s)ds} \,,\quad  \hbox{if}\quad t<a\\
\pi(a) \int_{\alpha }^{\beta }{\bar b}\left( c,t-a\right) {\bar \phi}\left( c-t + a\right) 
dc\,,\quad \hbox{if}\, , t\ge a\, ,\hbox{and}\,\,  t\le \alpha    \\
\end{array}
\right.
\label{3.3}
\end{equation}
where,
\begin{eqnarray*}
\pi(a)&=&e^{-\int_{0}^{a} \mu (s)ds }\\
{\bar b}(a,t)&=&b\left( a,t\right) \pi(a)\\
{\bar \phi}(a)&=&{\phi}(a)e^{\int_{0}^{a} \mu (s)ds }
\end{eqnarray*}
and the function $n_S(a,t)$ as in (\ref{3.3}) is locally integrable  ($n_S(a,t)\in L^1_{loc} (S)$). Writing $n_S(a,t)$ as $n_S(a,t;\phi)$, and defining the functions,
$$
\phi_{i+1}(a)=n_S(a,\alpha;\phi_{i})
$$
where $i\ge 0$ and $\phi_0(a)=\phi(a)$, we can  construct recursively the general solution of the McKendrick equation as $n(a,t)=n_S(a,t-q\alpha ;\phi_p)$, where $q=[t/\alpha]$, and 
$[t/\alpha]$ stands for the integer part of $(t/\alpha )$.
\end{theorem}

\begin{proof} 
To construct the solution of the McKendrick equation in the strip
$S$, the case  $(a=0,t=0)$ follows from the  boundary condition (\ref{3.1}). The
case $t<a$ and $t>0$ has been proved in (\ref{3.2}). We now  construct the solution for  $(a,t)\in  {\bar T}_1$, Figure \ref{fig1}, which implies that $a\ge 0$, $t\ge a$ and $t< a+\alpha$. Let  $(a^*,t^*)$ be the point such that,
$t^*> a^*$ and $t^*< a^*+\alpha $. By (\ref{1.5}), the characteristic line that passes by $(a^*,t^*)$   crosses the line $a=0$ at some time $t=t_1^*$, and 
$n(a^*,t^*) = n(0,t_1^*)e^{ - \int_{t_1^*}^{t^*} \mu (s-t_1^* )ds}$, where $t_1^* = t^* - a^*$. Imposing the boundary condition (\ref{3.1}) on this solution,  we obtain,  
$$
n(a^*,t^*) = e^{ - \int_{t_1^*}^{t^*} \mu (s-t_1^* )ds}\int_{\alpha }^{\beta }b\left( a,t_1^*\right) n\left( a,t_1^*\right) da\, .
$$ 
By hypothesis, as $t_1^*=t^* - a^*< \alpha$, we are in the conditions of solution (\ref{3.2}), 
and we have,
\begin{eqnarray*}
n(a^*,t^*) = e^{ - \int_{t_1^*}^{t^*} \mu (s-t_1^* )ds}\int_{\alpha }^{\beta }b\left( a,t_1^*\right) n\left( a,t_1^*\right) da\\
= \int_{\alpha }^{\beta }b\left( a,t_1^*\right) \phi\left( a-t_1^*\right) 
e^{- \int_{0}^{t_1^*} \mu (s+a-t_1^* )ds - \int_{t_1^*}^{t^*} \mu (s-t_1^* )ds} da\, .
\end{eqnarray*}
Therefore, we have shown that, for $t \ge a$ and  $t< a+\alpha $, that is, for $(a,t)\in {\bar T}_1$,
\begin{eqnarray}
n(a,t) &=& \int_{\alpha }^{\beta }b\left( c,t-a\right) \phi\left( c-t + a\right) 
e^{- \int_{0}^{t - a} \mu (s+c-t+a )ds - \int_{t - a}^{t} \mu (s-t+a )ds}dc \nonumber \\
&=& \int_{\alpha }^{\beta }b\left( c,t-a\right) \phi\left( c-t + a\right) 
e^{- \int_{c-t+a}^{c} \mu (s )ds - \int_{0}^{a} \mu (s)ds}dc \nonumber \\
&=& e^{-\int_{0}^{a} \mu (s)ds }\int_{\alpha }^{\beta }{\bar b}\left( c,t-a\right) \phi\left( c-t + a\right) 
e^{ \int_{0}^{c-t+a} \mu (s )ds }dc
\label{3.4}
\end{eqnarray}
where,
$$
{\bar b}(c,t-a)=b\left( c,t-a\right) e^{-\int_{0}^{c} \mu (s)ds }\, .
$$

For $a=0$ and $t=\alpha$,
\begin{equation}
n(0,\alpha) =\int_{\alpha }^{\beta }b\left( c,\alpha \right) n\left( c,\alpha\right) dc\, .
\label{3.5}
\end{equation}
As, for $c>\alpha$, $n\left( c,\alpha\right) \in {\bar T}_0$, we introduce (\ref{3.2}) into (\ref{3.5}), and we obtain  (\ref{3.3}). Therefore,  we have constructed the solution of the McKendrick equation in the strip $S$.

As $\phi (a) $ is locally integrable, the integrals in (\ref{3.3}) are  bounded for $t\le \alpha$ and $\phi_{1}(a)=n_S(a, \alpha ;\phi)\in L^1_{loc} ({\R}_+)$.  In order to construct the solution of the McKendrick equation for any $t>\alpha$, we  proceed by induction and take at each step the new initial condition $\phi_{i+1}(a)=n_S(a, \alpha ;\phi_{i})$ with $i\ge 0$. This justifies the formula $n(a,t)=n_S(a,t-q\alpha ;\phi_m)$, where $q=[t/\alpha]$. 
\end{proof}

We introduce now a technical simplification. Defining the new function $p(a,t)$ through,
$$
n(a,t)=e^{-\int_{0}^{a} \mu (s)ds }p(a,t)
$$
and introducing it in the McKendrick equation (\ref{1.1}), we obtain,
\begin{equation}
{\frac{\partial p(a,t)}{\partial t}} + {\frac{\partial p(a,t)}{\partial a}} 
=  0\, .
\label{3.r1}
\end{equation}
Assuming that the boundary condition for equation (\ref{3.r1}) is, 
$$
p(0,t)=\int_{\alpha}^{\beta} {\bar b}(a,t)p(a,t)\,da
$$
with 
${\bar b}(a,t)=b\left( a,t\right) e^{-\int_{0}^{a} \mu (s)ds }$, and the initial condition is, 
$$
\psi(a)=e^{\int_{0}^{a} \mu (s)ds }\phi(a)
$$
the Cauchy problem for the McKendrick equation (\ref{1.1}) is simply transformed into the Cauchy problem for the equation (\ref{3.r1}).  

 By hypothesis, for every $t$, ${\bar b}(a,t)$ is zero outside the interval $[\alpha,\beta]$. As, in general, the constant $\beta$ is not an integer multiple of $\alpha$, for every $t$, we can extend the function ${\bar b}(a,t)$ as a zero function 
in the interval $[\beta,\beta']$, where $\beta'=q\alpha$, $q\ge 2$ is an integer, and
$(q-1)\alpha <\beta$. Hence, without loss of generality,  
we assume  that $\beta=q\alpha$, where $q\ge 2$ is an integer. In the following, and to simplify the notation, we take this approach.

\begin{theorem} \label{T3.4} 
Let  $n(a,0)=\phi(a )\, (\in L^1_{loc} ({\R}_+))$ be a locally integrable initial condition for the McKendrick partial differential equation (\ref{1.1}), with $a\ge 0$, $t\ge 0$, and boundary condition (\ref{3.1}). Assume that $\mu (a)\in L^1_{loc} ({\R}_+)\cap C^{0}({\R}_+)$ is a non negative function. 
Assume further that, for every $t\ge 0$,  $b(a,t)$  is locally integrable and differentiable, and has compact support in the interval $[\alpha, \beta ]$, where $\alpha<\beta$, $\beta=q\alpha$, and $q\ge 2$ is an integer. Define the integer $m=[t/\alpha+1]$. Then,
the general solution of the Lotka renewal integral equation is determined recursively and is given by:

\noindent {a)} If, $0\le t\le \alpha$,
\begin{equation}
B_1(t) =\int_{\alpha }^{\beta } {\bar b}(c_1,t){\bar \phi}(c_1-t)\, dc_1
\label{3.rt1}
\end{equation}
where $B_1(t)=B(t)$, for $0\le t\le \alpha$.

\noindent {b)} If, $\alpha\le t\le  2\alpha$,
\begin{eqnarray}
B_{2}(t)&=&\int_{\alpha }^{t } {\bar b}(c_2,t)B_{1}(t-c_2)\, dc_2+\int_{t }^{\beta } {\bar b}(c_2,t){\bar \phi}(c_2-t)\, dc_2
\nonumber\\
&=&\int_{\alpha }^{t } {\bar b}(c_2,t)\int_{\alpha }^{\beta }{\bar b}(c_1,t-c_2){\bar \phi}(c_1+c_2-t)\, dc_1dc_2\nonumber\\
&+&\int_{t }^{\beta } {\bar b}(c_2,t){\bar \phi}(c_2-t)\, dc_2\, .\nonumber\\
\label{3.rt2}
\end{eqnarray}

\noindent {c)} If, $m\alpha\le t\le  (m+1)\alpha$, with $m>2$ and $q\ge m+1$,
\begin{eqnarray}
B_{m}(t)
&=&\int_{\alpha }^{t-(m-2)\alpha} {\bar b}(c_{m},t)B_{m-1}(t-c_{m})\, dc_{m}\nonumber\\
&+&\sum_{i=2}^{m-1}\int_{t-(m-i)\alpha }^{t-(m-i)\alpha+\alpha } {\bar b}(c_{m},t)B_{{m}-i}(t-c_{m})\, dc_{m}
\nonumber\\
&+&\int_{t }^{\beta} {\bar b}(c_{m},t){\bar \phi}(c_{m}-t)\, dc_{m} \, .
\label{3.rt3}
\end{eqnarray}
 
\noindent Moreover, $B(t)$ is continuous, and $n(a,t)$ is also continuous for $t\ge \beta$ and $a\in [0,\beta]$. For $t\ge a$, the general solution of the McKendrick equation is,
\begin{equation}
n(a,t) =\pi(a)B(t-a)
\label{3.rt5}
\end{equation}
where $\pi(a)$, ${\bar b}$ and ${\bar \phi}$ are as in Theorem \ref{T3.1}.
\end{theorem}

\begin{proof} 
As we have seen in the discussion preceding the theorem, it is sufficient to prove the Theorem for $\mu(a)=0$. By Theorem 
\ref{T3.1}, for $m=1$ and $t\le\alpha$, we have,
\begin{equation}
B(t):=B_1(t)=\int_{\alpha }^{\beta } {\bar b}(c_1,t)\phi(c_1-t)\, dc_1\, .
\label{3.r4}
\end{equation}
For $\mu \not=0$, we make the substitution 
$\phi\to{\bar \phi}$, as discussed above, and we obtain a).

We consider now the case $\alpha\le t\le2 \alpha$.  Then, 
\begin{eqnarray}
B_2(t)&=&\int_{\alpha }^{t } {\bar b}(c_2,t)n(c_2,t)\, dc_2+\int_{t }^{\beta} {\bar b}(c_2,t)\phi(c_2-t)\, dc_2\nonumber\\
&=&\int_{\alpha }^{t } {\bar b}(c_2,t)n(0,t-c_2)\, dc_2+\int_{t }^{\beta} {\bar b}(c_2,t)\phi(c_2-t)\, dc_2 \nonumber\\
&=&\int_{\alpha }^{t } {\bar b}(c_2,t)B_1(t-c_2)\, dc_2+\int_{t }^{\beta} {\bar b}(c_2,t)\phi(c_2-t)\, dc_2  
\label{3.r5}
\end{eqnarray}
where $B(t)=B_2(t)$ for $\alpha\le t\le2 \alpha$. Introducing 
(\ref{3.r4}) into (\ref{3.r5}), and with 
$\phi\to{\bar \phi}$, we obtain b).

With $m=[t/\alpha +1]\ge 2$, and assuming that $\beta\ge (m+1)\alpha$, due to the particular form of the characteristic curves in Figure \ref{fig1}, we obtain,
\begin{eqnarray}
B_{m}(t)
&=&\int_{\alpha }^{t-(m-2)\alpha} {\bar b}(c_{m},t)B_{m-1}(t-c_{m})\, dc_{m}\nonumber\\
&+&\sum_{i=2}^{m-1}\int_{t-(m-i)\alpha }^{t-(m-i)\alpha+\alpha } {\bar b}(c_{m},t)B_{{m}-i}(t-c_{m})\, dc_{m}
\nonumber\\
&+&\int_{t }^{\beta} {\bar b}(c_{m},t)\phi(c_{m}-t)\, dc_{m}
\label{3.r6}
\end{eqnarray}
where $m\alpha\le t\le (m+1) \alpha$ and $m\ge 2$, proving c). Clearly, $B(t)$ is continuous for $t\in [0,\beta]$, and $B(t)$ depends on $\phi(a)$ with $a\in [0,\beta]$. As $n(a,\beta) =\pi(a)B(\beta-a)$, due to the continuity of $B(t)$ in the 
interval $[0,t]$, $n(a,\beta)$ is now continuous for $a\in [0,\beta]$. As $B(t)$ remains continuous for $t>\beta$, then $n(a,t)$ is also continuous for $t\ge \beta$ and $a\in [0,\beta]$. 

Defining the new initial condition $\phi (a)=B(\beta -a)$, the solution
of the renewal equation is recursively constructed in the intervals $[\beta,2\beta]$,
$[2\beta,3\beta],\ldots$

To derive the general solution of the McKendrick equation, we use (\ref{1.9}), and we
obtain the  Theorem.
\end{proof}

Theorems \ref{T3.1} and \ref{T3.4} give the general solutions of the 
McKendrick equation and of the Lotka renewal equation as a function of the initial data and of the boundary condition. The fertility modulus is assumed time and age dependent, and the mortality modulus age dependent.

Dropping the time dependence on the fertility function $b$, the stability or instability of the solutions of the Lotka renewal integral equation and of the McKendrick equation follow
from the previous Theorems.

\begin{theorem}\label{T3.5} 
If the fertility function is time independent and
in the conditions of Theorem \ref{T3.4},  we have:

\noindent {a)} If  $r=1$, then, for every  $a\in [0,\beta ]$ and $t\ge a$,
$n(a,t)$ remains bounded as $t\rightarrow \infty $.

\noindent {b)} If  $r>1$, then, for every  $a\in [0,\beta ]$ and  $t\ge a$, $n(a,t)\rightarrow \infty $, as  $t\rightarrow \infty $.

\noindent {c)} If  $r<1$, then, for every  $a\in [0,\beta ]$ and  $t\ge a$, $n(a,t)\rightarrow 0 $, as  $t\rightarrow \infty $,\hfill\break
where $r$ is the Lotka growth rate of the population, as defined in (\ref{Lot}). In the limit $t\to \infty$, the solution of the Lotka renewal equation behaves as $B(t)\simeq r^{t/\beta}$.
\end{theorem}

\begin{proof}
By Theorem \ref{T3.4}, $B(t)$ is  continuous, and, as $n(a,\beta)=\pi(a)B(\beta-a)$, $n(a,\beta)$ is continuous in the closed interval $[0,\beta]$. Therefore, there exist
numbers $m$ and $M$ such that,
\begin{equation}
m\le n(a,\beta) \le M\, .
\label{3t5.1}
\end{equation}
By Theorem \ref{T3.1}, for $0\le t_1\le \alpha$, we have,
$$
n(0,t_1+\beta) =B(t_1+\beta)=
\int_{\alpha }^{\beta }{\bar b}\left( c\right) n\left( c-t_1 ,\beta\right) 
dc\, .
$$
As, $c-t_1\in [0,\beta]$ for $c\in [\alpha,\beta]$ and $0\le t_1\le \alpha$, by
(\ref{3t5.1}), we obtain,
\begin{equation}
mr\le B(t_1+\beta) \le Mr
\label{3t5.2}
\end{equation}
where, by (\ref{Lot}),
\[
r=\int_{\alpha}^{\beta }{\bar b}(c)dc=\int_{\alpha}^{\beta }b(c)e^{-\int_{0}^{c}\mu (s) ds}dc  
\]
is the Lotka growth rate, and $0<\alpha<\beta<\infty$.
Let us calculate now a bound for $n(a,\alpha+\beta)$, with $a\in [0,\beta]$. 
By (\ref{3t5.2}), and considering the case $\mu =0$, due to the particular form of the characteristic curves of Figure \ref{fig1}, we have,
\begin{eqnarray}
mr\le &n(a,\alpha+\beta)& \le M\qquad\hbox{if}\qquad r<1\nonumber\\
m\le &n(a,\alpha+\beta)& \le Mr\qquad\hbox{if}\qquad r>1 \, .
\label{3t5.3}
\end{eqnarray}
Then, by (\ref{3t5.3}),
$$
B(t_1+\alpha+\beta)=
\int_{\alpha }^{\beta }{\bar b}\left( c\right) n\left( c-t_1 ,\alpha+\beta\right) 
dc
$$
and,
\begin{eqnarray}
mr^2\le &B(t_1+\alpha+\beta)& \le Mr\qquad\hbox{if}\qquad r<1\nonumber\\
mr\le &B(t_1+\alpha+\beta)& \le Mr^2\qquad\hbox{if}\qquad r>1 
\label{3t5.4}
\end{eqnarray}
where $0\le t_1\le \alpha$. Repeating this procedure up to $q=\beta/\alpha$, and as,
for $\mu =0$,
$n(a,2\beta)=B(2\beta-a)$, the bound for $n(a,2\beta)$, with $a\in [0,\beta]$, must obey to the inequalities (\ref{3t5.2}),
(\ref{3t5.4}), etc., which gives,
\begin{equation}
mr\le n(a,2\beta) \le Mr\, .
\label{3t5.5}
\end{equation}
Comparing (\ref{3t5.1}) and (\ref{3t5.5}), the theorem follows by induction. In the asymptotic limit, the solution of the Lotka renewal equation behaves as $B(s\beta)\simeq r^{s}$, where $s$ is an integer. 
\end{proof}

With $b(c)=b\delta (a-\alpha)$ in Theorem \ref{T3.5}, we obtain Theorem \ref{T2.3}.

Let us see a simple example that shows that for a fertility function distributed along an age interval, we have damped growth  cycles with a period equal to the age $\alpha$. Suppose that $b(a)=b$ is constant in the  interval $[\alpha, \beta=2\alpha]$,  the mortality modulus is age independent, and the
initial population is constant in the interval $(0,\beta]$. Let us denote the solution of the renewal equation by $B_i(t)$ for $t\in [(i-1)\alpha, i\alpha]$, with $i\ge 1$. Then, by Theorem \ref{T3.4}, we have,
\begin{eqnarray}
B_1(t) &=&\phi e^{-\mu t}b\alpha \nonumber\\
B_2(t) &=&\phi e^{-\mu t} b(b\alpha (t-\alpha)+ (2\alpha-t)) \, .
\label{3.ex1}
\end{eqnarray}
By (\ref{3.ex1}) and (\ref{1.9}), we have,
\begin{equation}
n(a,2\alpha) =\pi(a)B(2\alpha-a)=\left\{ \begin{array}{l} 
\phi e^{-\mu 2\alpha} b(b\alpha (\alpha-a)+ a) \,,\quad  \hbox{if}\quad a\le\alpha \\
\phi e^{-\mu 2\alpha }b\alpha\,,\quad \hbox{if}\quad \alpha\le a\le 2\alpha\, .    \\
\end{array}
\right.
\label{3.ex2}
\end{equation}
With the new initial conditions $\phi(a)=n(a,2\alpha)$, by Theorem \ref{T3.4}a) and b), we obtain,
\begin{eqnarray}
B_3(t) &=&\phi e^{-\mu t}b^2(\alpha^2 +\frac{1}{2}(t-2\alpha)^2 (b\alpha-1)) \nonumber\\
B_4(t) &=&\phi e^{-\mu t}b^2\frac{1}{6}\left(bt^3 (b\alpha-1) -t^2(9b^2 \alpha^2-6b\alpha-3)\right.\nonumber\\
&+&\left.3t\alpha(9b^2 \alpha^2-b\alpha-8)-3\alpha^2(9b^2 \alpha^2+5b\alpha-16)\right) \, .
\label{3.ex3}
\end{eqnarray}

\begin{figure}[htbp]
\centerline{\includegraphics[width=3.23in]{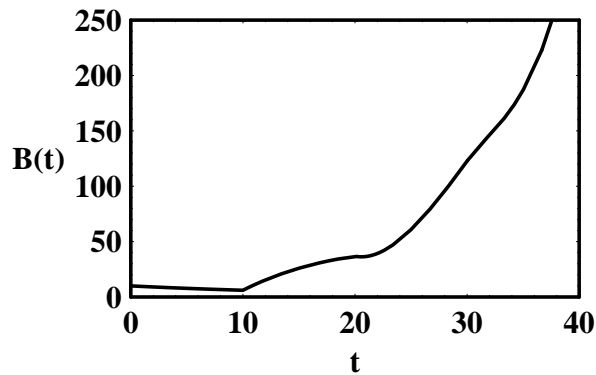}}
\caption{Time evolution of new-borns $B(t)$ for an initial uniform population, calculated from (\ref{3.ex1}) and (\ref{3.ex3}). The initial condition is $\phi(a)=1$, for $a\in (0,2\alpha=\beta]$, and the parameter values are:  $b =1.0$, $\alpha =10$ and $\mu=0.05$. }
\label{fig4}
\end{figure}

In this case we have, $B(0)=\phi b\alpha$, $B(\alpha)=\phi e^{-\mu \alpha}b\alpha$,
$B(2\alpha)=\phi e^{-2\mu \alpha}b^2\alpha^2$,  $B(3\alpha)=\phi e^{-3\mu \alpha}b^2\alpha^2(1+b\alpha)/2$ and $B(4\alpha)=\phi e^{-4\mu \alpha}b^3\alpha^3(5+b\alpha)/6$. In Figure \ref{fig4}, we show the time behaviour of
$B(t)$ for $\alpha =10$, $b=1$, $\phi=1$ and  mortality modulus $\mu=0.05$. In this case, the Lotka growth rate is $r=b (e^{-\mu \alpha} - e^{-2\mu \alpha} )/\mu=4.77$. 
From this example, we conclude that there are two time scales associated to the growth of new borns. The first time scale $\alpha $ is related with the transition from the solutions $B_i(t)$ to $B_{i+1}(t)$. The second time scale $\beta$ is associated with the   
Malthusian growth behaviour. From the above computed values of $B(i\alpha)$, with $i\ge 0$, we have $B(i\alpha)\simeq r^{[i\alpha/\beta]}$, in agreement with Theorem
\ref{T3.5}. In this example, there exists a damped modulation in the asymptotic time behaviour with period $\alpha$.

From Theorem \ref{T3.1}, we can derive the Leslie discrete model of population dynamics, \cite{b9}, and relate the stability
properties of both the continuous and discrete models.

\section{From the McKendrick model to the Leslie discrete model}\label{sML}

In this section, we assume that the fertility modulus is time independent.
Suppose
that the age axis is partitioned into intervals of length $\Delta a$ and the extreme of these intervals  are indexed by the integers $i\ge 0$: $a_i= i\Delta a$. Assume that $\Delta a$ is small, in the sense that $\Delta a<\alpha$. With $\Delta t=\Delta a$, by Theorem \ref{T3.1}, we have, for $i>1$,
$$
n_S(i\Delta a,\Delta t)=n_S(i\Delta a-\Delta a,0)e^{-\int_{i\Delta a-\Delta a}^{i\Delta a}\mu(s) ds}\, .
$$
Defining  $n_i^{\Delta t}=n_S(i\Delta a,\Delta t)$, we can write,
\begin{equation}
n_i^{\Delta t}=e^{-\int_{i\Delta a-\Delta a}^{i\Delta a}\mu(s) ds}n_{i-1}^{0}:=
p_{i-1}n_{i-1}^{0}= e^{-\Delta a \mu((i-1)\Delta a)} n_{i-1}^{0}\quad (i>1)
\label{3.6}
\end{equation}
where we have approximated the integral by its Riemann sum, $n_i^{\Delta t}$ is the density of individuals with age $a_i= i\Delta a$ at time $\Delta t$, and $p_{i-1}$ is the survival transition probability between age classes $i-1$ and $i$. For $i=1$, by Theorem \ref{T3.1},  
\begin{equation}
n_1^{\Delta t}=n_S(\Delta a,\Delta t)=\int_{\alpha }^{\beta }b\left( c\right) \phi\left( c\right) 
e^{-{ \int_{0}^{\Delta a} \mu (s )ds}}
dc\, .
\label{3.6b}
\end{equation}
Approximating the integrals by Riemann sums, we obtain,
\begin{equation}
n_1^{\Delta t}=e^{-\Delta a \mu(0)} \Delta a \sum_{m=q}^{s-1} b\left( m\Delta a\right) \phi\left( m\Delta a\right):= \sum_{m=q}^{s-1} e_m \phi\left( m\Delta a\right)
\label{3.7}
\end{equation}
where $q=[\alpha / \Delta a]$, $s=[\beta / \Delta a]$,
and $e_m$ are fertility coefficients. Writing (\ref{3.6}) and (\ref{3.7}) in matrix form, we obtain,
\begin{eqnarray}
\left( \begin{array}{c}
   {n_1^{\Delta t} }  \\ 
    \vdots   \\ 
   {n_{s - 1}^{\Delta t} }   
 \end{array} \right) 
= \left( \begin{array}{ccccccc}
   0 & 0 &  \cdots  & {e_q } &  \cdots  & {e_{s - 2} } & {e_{s - 1} }  \\
   {p_1 } & 0 &  \cdots  & 0 &  \cdots  & 0 & 0  \\
   0 & {p_2 } &  \cdots  & 0 &  \cdots  & 0 & 0  \\ 
    \vdots  &  \vdots  &  \cdots  &  \vdots  &  \vdots  &  \vdots  &  \vdots   \\ 
   0 & 0 &  \cdots  & 0 & 0 & {p_{s - 2} } & 0  
 \end{array} \right)\left( \begin{array}{c}
   {n_1^0 }  \\ 
    \vdots   \\ 
   {n_{s - 1}^0 }   
\end{array} \right)
\label{3.8}
\end{eqnarray}
which is the discrete Leslie model for age-structured populations, \cite{b9}. 
The parametric relations in (\ref{3.6}) and (\ref{3.7}) enable the comparison of 
both models, and in the limit $\Delta a\to 0$, the solutions of the McKendrick equation
converge to the solution of the Leslie model (\ref{3.8}). On the other hand, a fast way of computing numerically 
the solutions of the McKendrick equation is through (\ref{3.8}) with $\Delta a$ small. 

Given an initial distribution of
population numbers $(n_1^0,\ldots ,n_{s-1}^0)\not=0$, the asymptotic state of the linear system (\ref{3.8})
is zero or goes to infinity, depending on the magnitude of the
dominant eigenvalue of the matrix in (\ref{3.8}). 
If the dominant eigenvalue of the matrix in (\ref{3.8}) is $\lambda =1$,
bounded and non-zero population distributions are
obtained. 
The eigenvalues of the matrix in (\ref{3.8}) are determined by solving its characteristic polynomial equation, \cite{b3},
\begin{equation}
P(\lambda )=(-1)^{s-1} \left(\lambda^{s-1}-\sum_{i=q}^{s-1} \lambda^{s-1-i} e_i\prod_{j=2}^i p_{j-1}\right)=0 \, .
\label{3.9}
\end{equation}
As we have assumed that $\alpha < \beta$, and $\Delta a$ is sufficiently small, then
$e_{s-1}>0$ and $e_{s-2}>0$. Hence,  the matrix in (\ref{3.8}) is primitive and, by the Frobenius-Perron theorem,
the dominant root of the characteristic polynomial $P(\lambda )$ is simple, real and positive, \cite{b2}. Therefore,
the condition for the existence of a bounded and stable nonzero asymptotic solution
for Leslie map (\ref{3.8}) is, 
\begin{equation}
G(\Delta a):=\sum_{i=q}^{s-1}  e_i\prod_{j=2}^i p_{j-1}=
e_qp_1\ldots p_{q-1}+\cdots+e_{s-1}p_1\ldots p_{s-2}=1
\label{3.10}
\end{equation}
where the parameter $G(\Delta a)$ is the inherent net reproductive number of the population 
associated with the Leslie model (\ref{3.8}), \cite{b2}. Introducing the parameters $p_i$ and
$e_i$ as defined in (\ref{3.6}) and (\ref{3.7}) into (\ref{3.10}), we obtain,
\begin{equation} 
\begin{array}{lcl}
G(\Delta a)&=&e^{-\Delta a \mu(0)}  \Delta a \displaystyle \sum_{i=q}^{s-1} b\left( i\Delta a\right)\prod_{j=2}^i e^{-\Delta a \mu((j-1)\Delta a)}  \\[8pt]
&=&\Delta a \displaystyle \sum_{i=q}^{s-1} b\left( i\Delta a\right)e^{
-\Delta a \sum_{j=0}^{i-1}\mu((j)\Delta a)}\, .
\end{array}
\label{3.11}
\end{equation}
In the limit $\Delta a\to 0$, passing from Riemann sums to integrals, and
as $q=[\alpha / \Delta a]$ and $s=[\beta / \Delta a]$, we obtain for the inherent net reproductive number of the population,
\begin{equation}
\lim_{\Delta a\to 0}G(\Delta a)=\int_{\alpha}^{\beta }b(c)e^{-\int_{0}^{c}\mu (s) ds}dc =r
\label{3.12}
\end{equation}
which equals the Lotka growth rate (\ref{Lot}).
Therefore, in the limit $\Delta a\to 0$, the Lotka growth rate and the inherent net reproductive number of the population are the same quantities.

In the case of one fertile age class, the only
non-zero fertility coefficients in (\ref{3.8}) is $e_{s-1}$, and the matrix in (\ref{3.8}) is
non primitive but irreducible. In this case, by (\ref{3.9}), we have,
$$
P(\lambda )=(-1)^{s-1} \left(\lambda^{s-1}- e_{s-1}\prod_{j=2}^{s-1} p_{j-1}\right)=0 
$$
and the $s-1$ roots  of the Leslie matrix are within  the circle with radius, 
$$
\left(e_{s-1}\prod_{j=2}^{s-1} p_{j-1}\right)^{1/(s-1)}\, .
$$
Therefore, when the fertility modulus is concentrated at a point,  the ratio $n(a,t)/N_M(t)$ is an oscillatory function of time, as it has been shown in Theorem 
\ref{T2.6}, and $N_M(t)$ is the Malthusian growth function (\ref{2.10}). If the fertility modulus shows dispersion around some age class, the matrix in (\ref{3.8}) becomes  primitive. In this case, for $r>1$ but close to $1$,  the complex eigenvalues of the matrix in (\ref{3.8}) are in the interior of the circle with radius $r$ and, asymptotically in time,  oscillations dye out. However, as we have seen previously in (\ref{3.ex2}), the persistence or the damping of the amplitude of the oscillations  depend  on the initial distribution of the population.

\begin{figure}[htbp]
\centerline{\includegraphics[width=13cm]{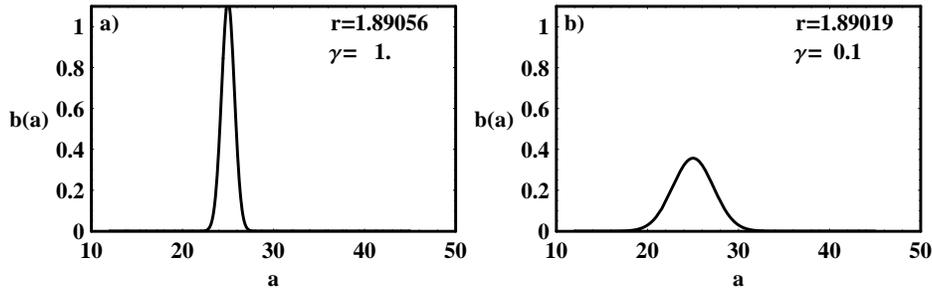}}
\caption{Fertility modulus (\ref{3.13}) as a function of the dispersion parameter $\gamma$. We show  the corresponding Lotka growth rates $r$  calculated from (\ref{3.12}), for the mortality modulus $\mu (a)=0.001+0.0001 a$, dispersion $\gamma $, and  parameter values: $\alpha =12$, $\alpha_0 =25$, $\beta =45$  and $b_1=2$.}
\label{fig5}
\end{figure}

To  study the fertility dispersive behaviour when we pass from one reproductive age class to several reproductive age classes, we introduce the Gaussian shaped fertility modulus,
\begin{equation}
b(a)=b_1\sqrt{\gamma \over \pi} e^{-\gamma (a-\alpha_0)^2}\quad \hbox{with} \quad a\in[\alpha,\beta]
\label{3.13}
\end{equation}
where $b_1$, $\alpha_0$ and $\gamma $ are parameters. If $\gamma \to 0$, the fertility of the population 
is distributed among several age classes. If $\gamma \to \infty$, the fertility is concentrated at the age class $a=\alpha_0$, in the sense that, $\lim_{\gamma \to \infty} \int_{-\infty}^{+\infty}\psi(a)b(a) da=b_1 \psi(\alpha_0)$, where $\psi(a)\in {\D}({\R}_+)$.
In Figure \ref{fig5}, we show the behaviour of the fertility modulus (\ref{3.13})  as a function of $a$ for several values of the dispersion parameter and consequently different Lotka growth rates.

\begin{figure}[htbp]
\centerline{\includegraphics[width=13cm]{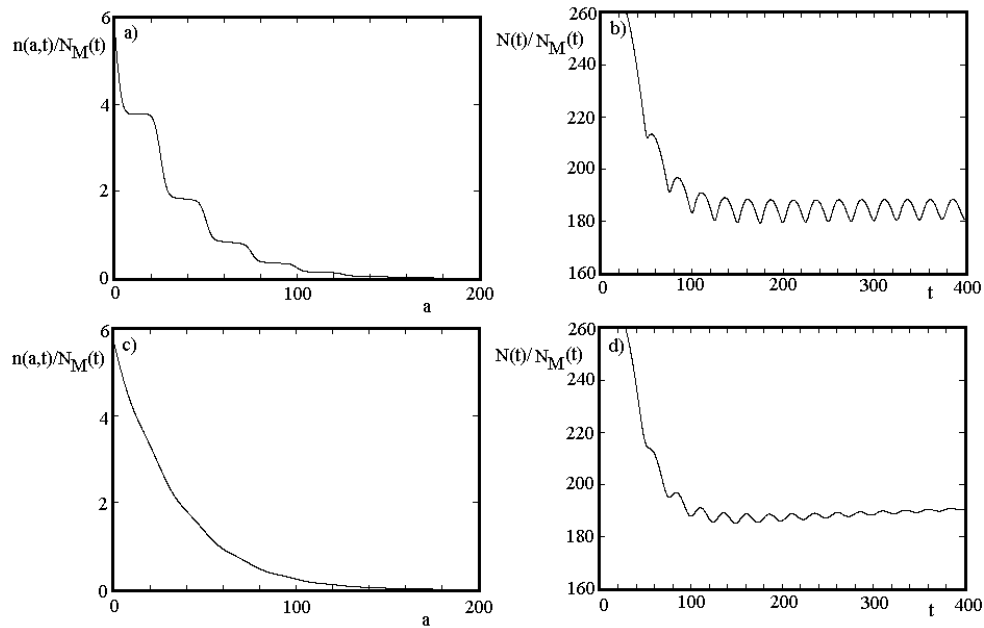}}
\caption{Ratios $n(a,t)/N_M(t)$ and $N(t)/N_M(t)$ for McKendrick equation with boundary condition (\ref{3.1}), calculated from initial data $\phi(a) = 2$,
for $a\le 100$, and $\phi(a) = 0$, for $a> 100$. The Malthusian growth function is
$N_M(t)= r^{(t-\alpha_0)/\alpha_0}$. In a) and c) the ratio of age distributions has been calculated at the  time $t=400$. a) and b) correspond to the fertility function of Figure \ref{fig5}a), and c) and d) correspond to the fertility function of Figure \ref{fig5}b). As it has been shown, small dispersion in the fertility modulus imply population oscillations. If fertility is not concentrated in one single age class, in the limit $t\to \infty$, oscillatory behaviour dies out.}
\label{fig6}
\end{figure}

In Figure \ref{fig6}, we show the ratios $n(a,t)/N_M(t)$ and $N(t)/N_M(t)$, where 
$N_M(t)= r^{(t-\alpha_0)/\alpha_0}$ is the Malthusian growth function as in (\ref{2.10}),
but with $r$ given by (\ref{3.12}). As it is expected, for small
dispersion around the class of maximal fertility, oscillations persist for a long transient time, but as we increase dispersion, their amplitude decreases in time. In the limit
$t\to \infty $, the oscillations disappear. Therefore, in the asymptotic limit, the solution of the McKendrick equation have the form $n(a,t)\approx c_1 r^{c_2t}$, where $c_1$ and $c_2$ are constants and $r$ is given by (\ref{3.12}).

\section{Predictions from demography data}\label{sDP}

For a population with $n$  age classes corresponding to reproductive ages $\alpha_i$, with 
$i=1,\ldots ,n$, we  define the 
fertility function,
  \begin{equation}
b(a) = \sum_{i=1}^nb_i \delta(a-\alpha_i)
\label{3.14}
\end{equation}
and the boundary condition for the McKendrick equation is now,
\begin{equation}
n(0,t) = \sum_{i=1}^nb_i n(\alpha_i ,t)\, .
\label{3.15}
\end{equation}
By (\ref{Lot}), the growth rate of the population is,
\begin{equation}
r=\int_{\alpha}^{\beta }b(a)e^{-\int_{0}^{a}\mu (s) ds}da=\sum_{i=1}^nb_i e^{-\int_{0}^{\alpha_i}\mu (s) ds}:=\sum_{i=1}^n r_i\, .
\label{3.16}
\end{equation}
In this case, the asymptotic behaviour of the solution of
the McKendrick equation  is determined by
the Lotka growth rate of each cohort.

Due to the linearity of the McKendrick equation (\ref{1.1}), and the boundary condition 
(\ref{1.2}), the general solution is now the sum of the solutions for each $r_i$ as in Theorem \ref{T2.3} for time independent fertility function. In this case, each individual solution will 
have an exponential pattern of growth, modulated by a periodic function with period $\alpha_i$. Therefore, the pattern of growth of a population with several fertile age classes is almost periodic with several periodicities or frequencies. If the dispersion of the fertility around the maximal fertility age class is large, the pattern of growth is almost periodic in time. If, ~in addition, we consider time variations in the fertility function and mortality modulus, the pattern of growth can show strong fluctuations deviating from the periodic or exponential growth. This introduces a higher degree of unpredictability for the long time behaviour of population growth.

One of the important aspects of these results is that the period of oscillations
of population cycles are of the order of the age of one generation, as observed in human populations, \cite{b14}. In the Easterlin model, the period of the cycles are of the order of two generations, \cite{key}.

\section{Conclusions}\label{s4}

We have obtained the weak solutions in the sense of distributions of the McKendrick equation and of the Lotka renewal integral equation. We have assumed an age and time dependent fertility modulus and an age dependent mortality modulus.

  If $\mu (a)$ is the age dependent death rate of a species, and $b(a)$ is the age dependent fertility modulus, the Lotka growth rate is defined by,
\begin{equation}
r=\int_{\alpha}^{\beta }b(c)e^{-\int_{0}^{c}\mu (s) ds}dc
\label{4.1}
\end{equation}
where $\alpha $ and $\beta $ are the first and last reproductive age classes. The Lotka growth rate $r$ 
determines the stability of the solutions of the McKendrick equation, in the sense that,
asymptotically in time, we have exponential growth if $r>1$, and extinction if $r<1$. If $r=1$, we have an asymptotically stable population and the equilibrium solution is
$n(a)=n_0\exp\left({ - \int_{0}^{a} \mu (s)ds}\right)$, where $n_0$ is a constant.

From the demography data point of view, (\ref{4.1}) implies that  measured fertility numbers by age class are given by
$$
b(a)e^{-\int_{0}^{a}\mu (s) ds}\, ,
$$
partially justifying the general non symmetric shape of the fertility curves in human po\-pu\-la\-tions, \cite{b8}.

One of the important features of the solutions of the McKendrick equation relies on the existence of a natural time scale determined by the age of the first fertile age class. In this case, a constant initial population will present cycling behaviour
in the patterns of growth. The period of these cycles is equal to the age of the first fertile age class. This contrasts with the prediction of Easterlin where the period equals
twice the mean age of one generation,  \cite{key}. Observations in human populations corroborate this result, \cite{b14}. These oscillations  are in general damped, and the damping is proportional to the magnitude of the dispersion of the fertility function around the most fertile age class. These demography oscillations have been observed in human populations, \cite{b14}, and  in bacterial growth in batch cultures, \cite{b13}.

On the other hand, it has been shown that the general solutions of the McKendrick equation and of the Lotka
renewal equation retain the memory from initial data.

In the limiting case of a population with only one reproductive age class, the 
modulation of the growth curve is always periodic and has no damping. The amplitude of oscillations increases if the mortality modulus decreases, and the limit of pure Malthusian growth
is obtained if the mortality modulus goes to zero. The population retains the memory from the initial data through the amplitude of oscillations of the Easterlin cycles.

It follows also from this approach that the McKendrick equation is associated with a small time scale, when we compare it with the time scale associated with the Malthusian growth law. These two time scales describe the evolution of a population at two different long range levels. In other words, the Malthusian growth law can be obtained from the McKendrick equation if
the mortality is zero, the fertility is concentrated at one age, and the initial population is uniform along the age variable.
Denoting by $\alpha_0 $ the age of the unique fertile age class, the population growth measured in time steps of $\alpha_0 $ is exponential or Malthusian. That is, the asymptotic solution of the McKendrick equation behaves as,
$$
n(a,t=k\alpha_0)= n(a,\alpha_0)r^{(k\alpha_0-\alpha_0)/\alpha_0}=
n(a,\alpha_0)r^{(k-1)}
$$
for $k= 1,2,\ldots $

We have explicitly derived the Leslie time discrete model of population dynamics from the general solutions of the McKendrick equation, enabling a direct calibration of the solutions of the McKendrick equation with  population data. We have shown that, if the length of the age classes goes to zero, the inherent net reproductive number of a population, the growth parameter 
of the Leslie model, converges to the Lotka growth rate.

\section*{Acknowledgments}
This work has been partially supported by the POCTI Project /FIS/13161/1998 (Portugal).

\end{document}